\documentclass[amsmath,amssymb,nofootinbib,twocolumn,aps,prl]{revtex4-2}
\usepackage{graphicx}
\usepackage{amsmath}
\usepackage{xcolor}

\begin{document}
\title{Few-body hydrodynamic interactions probed by optical trap pulling experiment}
\author{Julian Lee}
\affiliation{Department of Bioinformatics and Life Science, Soongsil University, Seoul 06978, Korea}

\author{Kyle Cotter, Ibrahim Elsadek}
\affiliation{Department of Physics and Astronomy, Michigan State University, East Lansing, Michigan 48824, USA}

\author{Matthew J Comstock}
\email{mjcomsto@msu.edu}
\affiliation{Department of Physics and Astronomy, Michigan State University, East Lansing, Michigan 48824, USA}

\author{Steve Press\'e}
\email{spresse@asu.edu}
\affiliation{Department of Physics, Arizona State University, Tempe, Arizona 85287, USA}

\date{\today}
\begin{abstract}
{We study the hydrodynamic coupling of neighboring micro-beads placed in a dual optical trap setup allowing us to precisely control the degree of coupling and directly measure time-dependent trajectories of the entrained beads. Average experimental trajectories of a probe bead entrained by the motion of a neighboring scan bead are compared with theoretical computation, illustrating the role of viscous coupling and setting timescales for probe bead relaxation. The findings provide direct experimental corroborations of hydrodynamic coupling at larger, micron spatial scales and millisecond timescales, of relevance to hydrodynamic-assisted colloidal assembly as well as improving the resolution of optical tweezers. We repeat the experiments for three bead setups. 
} 
\end{abstract}
\maketitle
{\it Introduction}.---Forces between solid bodies immersed in fluid are mediated through hydrodynamic interactions~\cite{micro} and the effects of such interactions abound across physical~\cite{achiral, snijkers} and biological systems~\cite{cy1}.
Though theories of hydrodynamic interaction have been constructed~\cite{nsp1,nsp2,sk87,nsp3,fh1,pad,jcp7,jcp14,3sp,jcp8,jcp15,hyd6,ourjcp}, their experimental corroborations have relied on indirect methods such as the analysis of collective motions of colloidal particles~\cite{add1,add2,add3, ladd95,val07,nov10, val18}.

Similarly, hydrodynamic coupling effects on particle behavior have been deduced from the analysis of thermal fluctuations of diffusing colloidal particles~\cite{lin02,cui02,cui04,misi15,val18,zheng22} or two micrometer-sized objects in stationary optical traps~\cite{dur00,add3,mar06,leo11}, limiting the comparison to theoretical frameworks at spatiotemporal scales of $\mu {\rm s}$ and ${\rm nm}$. 

By contrast, here we study the effect of a neighboring ``scan bead" made to move with respect to a probe bead immobilized in an optical trap. In doing so, we capture the onset of viscous coupling, its development, resulting probe bead entrainment, and subsequent relaxation manifest at larger spatial and longer temporal scales.

These scales are relevant to the self-assembly of colloidal spheres, naturally mediated through hydrodynamic interactions~\cite{achiral,assembly}. Of equal, and practical, importance is understanding hydrodynamic coupling of beads held at short distances, as we explore here, relevant in pushing tweezer experiments to higher resolution~\cite{opttwrev}. By contrast, despite their resolution advantage, short distances are typically avoided to mitigate the convolution of signal with bead hydrodynamics~\cite{opttwrev}. 

In this Letter, we probe hydrodynamic coupling by directly measuring the average trajectory of micron-sized bead in a stationary optical trap (probe bead), in the presence of one or two scan beads in the vicinity pulled using moving optical traps. This dual trap setup allows us to collect repeated statistics in a way that would not otherwise be possible if we were to wait for the spontaneous alignment of freely diffusing beads in solution, as with experiments on colloidal suspensions. Good statistics are especially critical in allowing us to directly compare theoretical probe bead trajectories while overcoming severe thermal agitation of beads otherwise corrupting any individual trajectory.  

{\it Theory}.---
The equation of motion for hard spheres moving in viscous fluid has previously been derived by generalizing the equation for one sphere in a fluid~\cite{micro,kynch,nsp1,nsp2,sk87,nsp3,3sp,2sp,2spnew}. The timescales probed by the current experiment are of order ${\rm ms}$, on which micro-bead motions are overdamped. In such a regime, bead dynamics are captured by
\begin{equation}
\frac{d {\bf x}}{d t} = {\boldsymbol \mu}{\bf F}    
\end{equation}
where ${\bf x}$ and ${\bf F}$ are $3N$ dimensional vectors describing the coordinates of $N$ beads and the external forces acting on the beads, respectively. Here  ${\boldsymbol \mu}$ is a rank-two tensor of size $3N$ called the mobility tensor. 
The mobility tensor ${\boldsymbol \mu}$ is obtained by computing the fluid flow in the presence of a sphere of radius $R$ with no-slip boundary conditions using a low Reynolds number Navier-Stokes equation where both convective and time-derivative terms are neglected~\cite{LL}. The mobility tensor is normally expressed as a power series in $\epsilon \equiv r/R$, where $r$ is the radius of each sphere and $R$ is the distance between them~\cite{micro,kynch,nsp1,nsp2}. 
Defining $\tilde {\boldsymbol \mu}_{ij} \equiv 6 \pi \eta {\bf \mu}_{ij}$, where $i,j=A,B,C$ are the indices for the spheres with corresponding radii $a$,$b$, and $c$, we have
\begin{eqnarray}
\tilde {\boldsymbol \mu}_{AA}  &=& \frac{1}{a} - \frac{15 b^3}{4 R_{AB}^4} {\bf \hat e}_{AB} {\bf \hat e}_{AB}  - \frac{15 c^3}{4 R_{AB}^4} {\bf \hat e}_{AC} {\bf \hat e}_{AC} + O(\epsilon^{6}),  \nonumber\\
\tilde {\boldsymbol \mu}_{AB}  &=& \frac{3}{4R_{AB}}({\bf I}+{\bf \hat e}_{AB}{\bf \hat  e}_{AB}) + \frac{a^2+b^2}{4 R_{AB}^3}({\bf I}-3 {\bf \hat e}_{AB}{\bf \hat  e}_{AB}) \nonumber\\
&&- \frac{15 c^3}{8 R_{AC}^2 R_{CB}^2} (1-3 ({\bf \hat e}_{AC}\cdot {\bf \hat  e}_{CB})^2 ) {\bf \hat e}_{AB}{\bf \hat  e}_{CB}\nonumber\\
&&+O(\epsilon^6)\label{3bdeq}
\end{eqnarray}
where ${\bf \hat e}_{ij}$ is the unit vector connecting the two spheres $i$ and $j$, and $I$ is the unit tensor. The expression for $\tilde \mu_{AC}$ is obtained from that for $\tilde \mu_{AB}$ by exchanging the labels $B$ and $C$. We explicitly verified that the addition of truncated terms does not result in any visible change in the computed graph of the bead motions. It is straightforward to obtain the mobility for the two beads, by taking the limit of $R_{AC}, R_{BC} \to \infty$ in Eq.(\ref{3bdeq})~(See supplementary material for details).

\begin{figure}
\includegraphics[width=\linewidth]{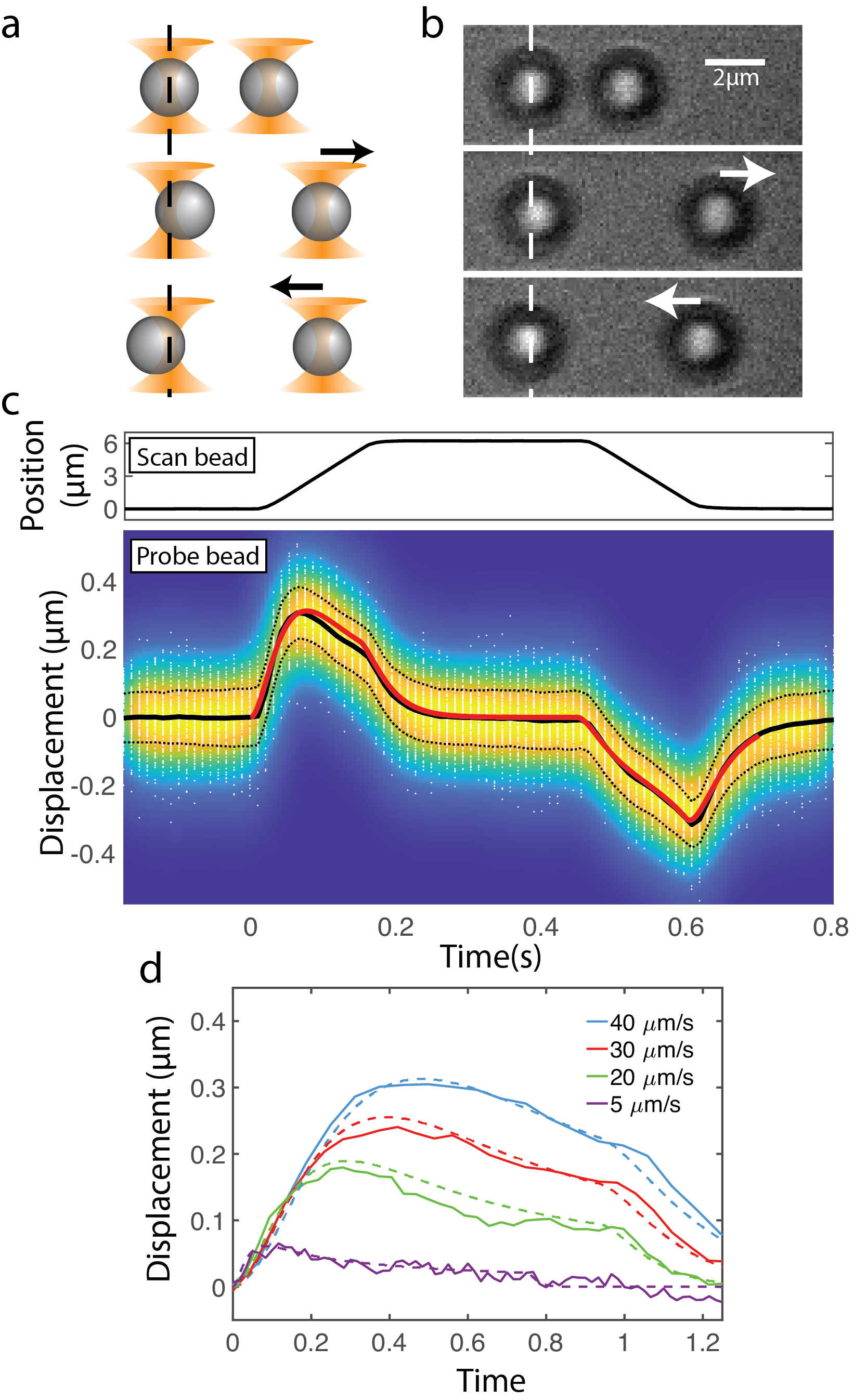}
\caption{Two bead linear scan data. a) Cartoon illustrating the general two bead setup. Optical traps (orange cones) hold microbeads (gray spheres). A trap moves a scan bead away (middle frame) and then back towards (bottom frame) a probe bead held in a fixed trap. b) Images of linear scanning experiment as in a).  Dashed line indicates the fixed trap position and the probe bead zero displacement as a guide to the eye. White arrows indicate the scan direction. c) Linear scan data. Upper: Average position of $2.16 \mu m$ scan bead. Lower: Displacement of $2.16 \mu m$ probe bead. Black line is the displacement trajectory averaged over 19 unique bead sets (nominally identical but distinct pairs of scan and probe beads) with 100 repeated scans each. White dots are the actual bead displacements for each scan and the background color map the displacement distribution for one example bead set. The black dashed lines represent the RMS of all individual bead sets combined in quadrature. The red line is the simulated bead trajectory. d) Probe bead displacement (solid line) and corresponding simulated trajectories (dashed lines) for varying scan speeds. The time axis is normalized across all scan speeds by dividing time by the total scan time for the respective scan speed ({\it i.e}, t = 0 is the start of the scan and t = 1 is the end of the scan). The same set of beads are used for all curves and each curve results from the average of 100 individual scans.}
\label{linear2} 
\end{figure}

\begin{figure}
\includegraphics[width=\linewidth]{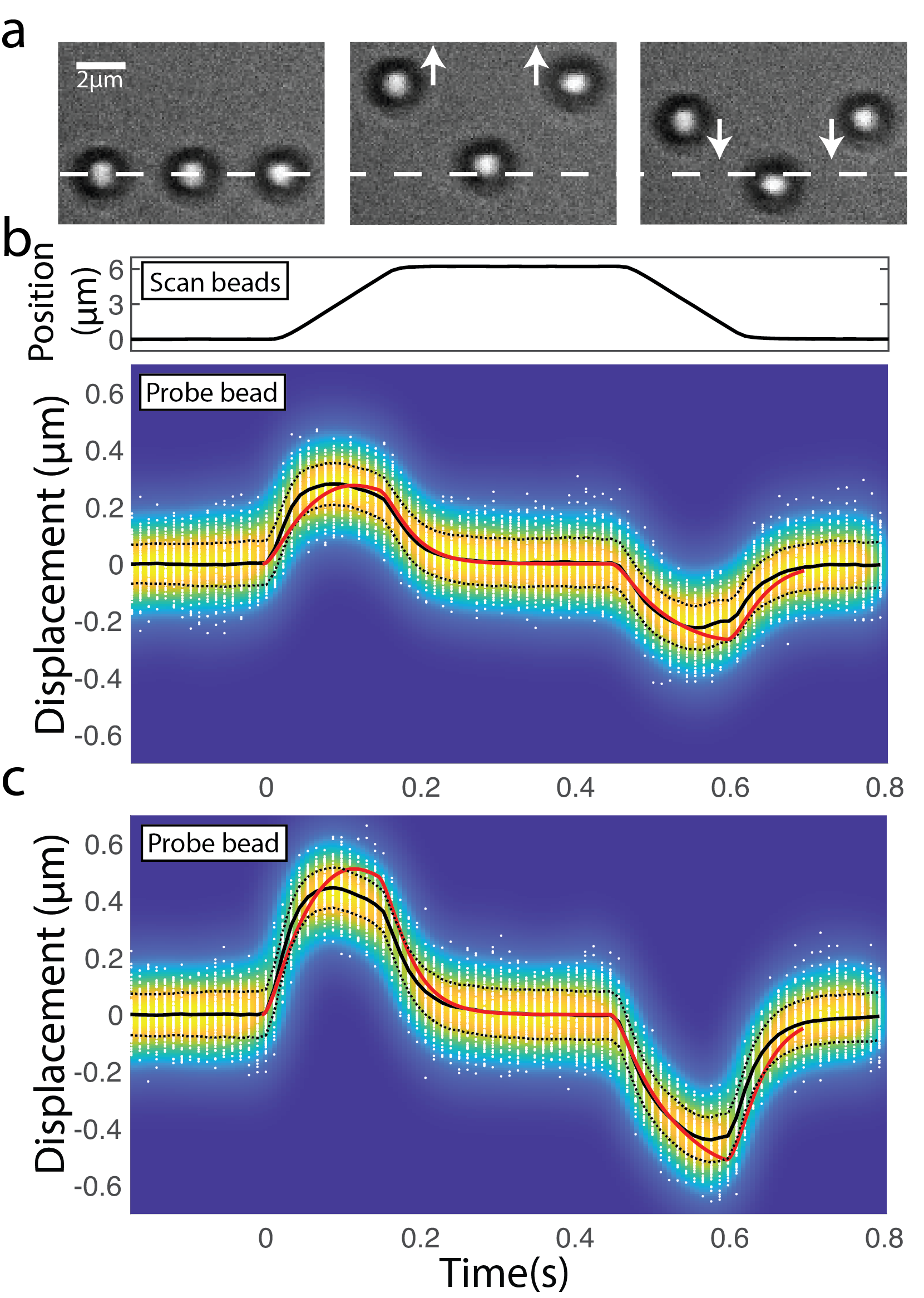}
\caption{Two- and three-bead shear scan data. a) Example images of a three-bead shear mode scanning experiment. For a two-bead shear scanning experiment, the right side scanning bead would be absent.  The dashed line indicates both the probe bead's fixed trap position as well as the initial scan bead trap positions. White arrows indicate the scan directions. b) and c) illustrate shear scan data. b) Upper: Average position of 2.16 $\mu$m scan bead. b) Lower and c): Displacement of 2.16 $\mu$m probe bead for two- and three-bead shear scan experiments, respectively. The black line is the displacement trajectory averaged over 11 and 19 unique bead sets for b) and c) respectively with 100 repeated scans each. White dots are the actual bead displacements for each scan and the background color map the displacement distribution for one example bead set. The black dashed lines represent the RMS of all individual bead sets combined in quadrature. The red line is the simulated bead trajectory.}
\label{shear23} 
\end{figure}

{\it Experimental Setup}.---
The hydrodynamic coupling between individual microspheres (2.16~$\mu$m  diameter polystyrene beads) was investigated by measuring the motion of a ``probe bead" held in a fixed position trap adjacent to a ``scan bead" quickly moved away and then back towards the probe bead ({\it e.g.}, as in Fig. \ref{linear2}a). A home-built high-resolution optical tweezers instrument was used, constructed and operated generally as described in \cite{neuman_block} and more specifically as in \cite{Yadav2022}, and shown in Supp. Fig. 4. Three optical traps were formed: two scanning traps flanking a fixed position probe bead trap. Traps were precisely manipulated via timesharing a single 976 nm laser using a two-axis acousto optic deflector (IntraAction DTD-274HD6C) controlled via a direct digital RF synthesis and field programmable gate arry (FPGA) method. Each trap in sequence was on for $5 \mu$ s duration. For all experiments, two scanning traps begin flanking the probe bead fixed trap with all three traps arranged in a line, and a scan is initiated by moving the two scanning traps in mirror trajectories away from then back towards the probe bead trap. For the linear scans the two scanning traps remain in a line with the probe bead trap (as in Fig. \ref{linear2}a) while for the shear scans the two traps move orthogonal to the initial alignment (as in Fig. \ref{shear23}a). For two-bead experiments, only a single scan trap is occupied by a bead (\textit{i.e.}, the other scan trap is empty), while for three-bead experiments both scan traps are occupied by beads. Bead positions were measured via imaging using CMOS camera (Thorlabs CS165MU) at 92 frames/second (10.9 ms per frame) which was synchronized with the scans via a signal from the FPGA. The probe bead trap stiffness (nominally $8.5 pN/\mu m$) was minimized to maximize the measured deflection of the probe bead in response to scan bead hydrodynamic coupling. The trap stiffness for each individual bead was calibrated via the standard equipartition method ({\it i.e.}, using the root-mean-square (RMS) of the bead Brownian motion; Supp. Fig. 5) and confirmed by measuring the bead's viscous relaxation time constant within the trap (Supp. Fig. 5). Control measurements were performed with alternate bead and scan configurations to confirm that the measured displacements of the probe bead are due to entrainment with the scan beads rather than spurious trap interactions (see Supp. Figs. 6 and 7).  

{\it Results}.---
We were able to directly detect and record trajectories of the probe bead in response to the fluid flow generated by the scan bead. The simplest perturbation to the probe bead is produced by an experimental configuration where a single bead is scanned in 1D along the line connecting probe and scan beads (subsequently referred to as ``two-bead linear scans"). Figure \ref{linear2}b shows a sequence of images from an example of an experiment with a probe bead in a fixed position trap on the left and a scan bead in a scanning trap on the right. The beads are assumed identical (as we later average over multiple experimental repetitions involving different probe and scan beads). The upper image shows the initial positions of the trapped beads before initiating the scan, with the closest distance between the probe and scan surfaces of $1 \mu $m. The middle image shows a frame of the scan where the scan bead is moving quickly ($40 \mu$m/s) away from the probe bead and the probe bead can be seen clearly displaced from the fixed trap minimum toward the scanning bead ($\approx 300$ nm displacement). The lower image shows a frame during the reversal of the scan, where the scan bead is now moving quickly (same speed) towards the probe bead and the probe bead can be seen clearly displaced from the fixed trap pushed away from the approaching scan bead ($\approx 300$ nm displacement).

The complete time trajectory of the entrainment of the probe bead in response to scan bead motion is shown in Fig. \ref{linear2}c. The solid black line shows the average probe bead displacement in the horizontal direction ({\it i.e.}, along the scan direction) vs time averaged over 19 bead pair replicates, where each replicate itself is an average of 100 repeated scans. At $t = 0$ the scan bead starts moving away at constant speed of $40\ \mu {\rm m/s}$ until halting $6\ \mu {\rm m}$ away at $t = 0.15\ {\rm s}$. In response, the probe bead is entrained by the scan bead (positive displacement), as the fluid flow induced by the moving scan bead flows past the probe bead and exerts a force on it in turn. The probe bead is much slower than the scanning bead, its  average speed being $\sim 6\ \mu {\rm m/s}$ from $t=0$ until it reaches the maximum displacement of $\sim 0.3\  \mu {\rm m}$ around $t \sim 0.05$\ s. As such, hydrodynamic interactions decrease with time due to the growing inter-bead distance, whereas the opposing force of the trap increases. The trap force takes over the hydrodynamic interaction after $t \sim 0.05$\ s, leading to a gradual relaxation of the probe bead back towards the center of the fixed trap for the remaining 2/3 of the scan duration. The probe bead displacement trajectory and maxima depended only weakly on the exact initial probe-scan bead separation (see Supp. Fig. 8). Upon halting the scan at $t = 0.15$ s, the probe bead rapidly relaxes to the center of the trap following a simple exponential decay. After a $0.3$ s pause (sufficient to ensure the probe bead is completely relaxed), the scan reverses and the scan bead moves back towards the probe bead at a constant speed of $40 \mu {\rm m/s}$. The probe bead is now pushed away from the scan bead (negative displacement). Note that this probe bead motion is not the mirror of the prior positive displacement motion. Indeed, since the push at the second scan starts at larger inter-bead distance compared to the pull at the first scan, the hydrodynamic coupling is also weaker initially, and becomes large only at the later stage where the restoring force of the trap becomes large. Consequently, the distance from the trap center  increases slowly and monotonically until it reaches $\sim 0.3\ \mu$m in the negative direction at $0.6$ s, at which point the reverse scan stops, leading to a simple exponential decay in displacement afterward. After a $0.4$ s pause, the scan repeats for a total of 100 cycles. The Brownian fluctuations of the bead position are significant compared to the maximum mean displacement trajectories (RMS of probe bead fluctuations is 80 nm while the maximum displacement is 300 nm, as seen directly by comparing individual trajectories and the distribution of these from Fig. \ref{linear2}c). The theory, as described above, simulates the probe-bead-scan-bead hydrodynamic coupling  reproducing the predicted mean probe bead trajectory (in red) which 
easily fits within the noise of the Brownian fluctuations. 

Since the magnitude of the hydrodynamic interaction is directly proportional to the speed of the objects moving in the fluid~(see SI Eq.(9)), we expect that the inter-bead coupling to be weakened if we decrease the speed of the scan bead. The average trajectory of a single bead set for the linear scan is shown in Fig.~\ref{linear2}d, along with the computational result. We see indeed that the maximal displacement of the probe bead decreases with the speed of the scan bead, in accordance with the theoretical prediction. Note that since we are averaging over only 100 trajectories for a single bead set, the effect of the thermal noise is more prominent than the data shown in (b) and (c). 

We also performed ``shear" configuration experiments where two or three beads flanking the probe bead are scanned orthogonal to the initial line connecting the probe and scan beads. Figure \ref{shear23}a shows a sequence of images demonstrating the shear mode scan configuration for the case of two flanking scan beads. Nearly all scan parameters are identical to the linear scan mode parameters (Fig. \ref{linear2}): all beads are nominally identical, the initial trap positions are identical with the closest distance between the probe and scan bead surfaces 1 $\mu$m and the scanning trap speed is $40 \mu$m/s. The only difference is the direction of the scan (which is vertical rather than horizontal). The experiments were repeated with either one or two flanking beads.

Complete trajectories of the displacement of the probe bead in response to the shear mode scanning beads with either one or two scanning beads are shown in Fig. \ref{shear23} (b) and (c) respectively. The solid black lines show the average probe bead vertical displacement ({\it i.e.}, parallel to the direction of the scan) vs time averaged over many bead pair replicates, where each replicate is an average of 100 repeated scans of the same beads. At $t = 0$ the scanning bead starts moving away at a constant speed of $40 \mu $m/s until halting $6 \mu $m away at $t = 0.15$ s. As for the linear scans, the probe bead follows the scan beads. For two flanking scan beads scanning in the same direction (as shown) the probe bead motion is completely vertical whereas for a single scan bead there is a slight horizontal displacement following the scan bead. If two scanning beads scan in opposite directions ({\it i.e.}, one upwards and one downwards) no displacement of the probe bead is observed as expected (Supp Figs 6 and 7). Both the initial speed ($\approx 9.5\ \mu$m/s) and the maximum displacement ($0.45\ \mu m$) of the probe bead are about 50\% greater with two scan beads, compared to those with one scan bead ($\approx 6\ \mu$m/s and 0.28 $\mu$m). The probe bead is pulled toward the retreating scan bead during the initial scan (positive displacement). After the intermediate scan pause, the probe bead is pushed away from the approaching scan beads (negative displacement). Again, this trajectory following the pause and reversal of direction of the scan bead is not the simple reversal of the trajectory preceding the pause, for the same reasons as was the case of the linear scan. The relaxation of the probe bead at scan halts are exponential with identical time constants in all cases.  

It is substantially more difficult to simulate the fluid coupling effects for shear scan, because one has to model two-dimensional motions of interacting three bodies, by contrast to the linear scan that can be described by the one-dimensional motion of two bodies. For the 2D shear scans, in both cases of one and two scan beads, the theory as described above predicts the probe bead trajectories still within the uncertainty of the probe bead Brownian fluctuations, though the mean does not match as closely as for the linear case. This may be due to incomplete modeling of external torques acting on the beads by the optical trap, which may play an important role in 2D motion.

{\it Conclusions}.---
In this work, we studied two- and three-body hydrodynamic interactions by directly probing the trajectory of a microbead in a stationary optical trap hydrodynamically entrained by neighboring microbeads. By contrast to previous work where thermal fluctuation of beads were analyzed to obtain information on hydrodynamic interaction in $\mu {\rm s}$ timescale~\cite{add3}, we averaged out thermal fluctuation to obtain dynamic trajectory on longer, ${\rm ms}$, timescales. Furthermore, we probed three-body in addition to two-body interactions. The resulting data agree with theoretical predictions within Brownian noise. 

Since our raw data prior to averaging contains thermal noise, it would also be interesting to compare the properties of such noise to the theoretical predictions. The theory for moving beads inside viscous fluid supplemented by  thermal fluctuations is the object of future investigation.

What is more, we have limited ourselves to the effects of entrainment manifest from linear motion though rotational diffusion and transfer of rotational chirality are also of immediate interest to questions of colloidal self-assembly~\cite{achiral} and the subject of future study. 

Finally, the conclusions drawn here are relevant to optical tweezer experiments where beads are often limited to operating at far distance from one another. Such experiments are performed precisely to eliminate the effects of bead-bead coupling. Yet for short bead-bead distances, as the spring constant of the tether typically goes inversely with the end-to-end tether distance~\cite{opttwrev}, models to account for coupling under this scenario are a key step toward improving the tether's sensitivity to the dynamics of the single molecule.  




\section{Acknowledgement} 
JL was supported by the National Research Foundation of Korea, funded by the Ministry of Science and ICT (NRF-2020R1A2C1005956).
KC, IE, and MJC were supported by NSF (MCB-1919439). SP was supported by NIH NIGMS (R01GM130745) and NIH NIGMS (R01GM134426). 


\bibliography{spheres}

\end{document}


\title{Supplementary Information}
\maketitle
\section{Details on theoretical analysis}

The dynamics of viscous liquid, such as the water at micrometer scale, is described by the linear Navier-Stokes equation
~\cite{LL}:
\begin{equation}
	{\rho}\frac{\partial \bf v}{\partial t} = - {\nabla p} + \eta \nabla^2 {\bf v}.\label{ulns}
\end{equation}
along with the condition of incompressibility
\begin{equation}
\rho={\rm constant} \leftrightarrow   
 \nabla \cdot {\bf v} = 0,\label{inc}
\end{equation}
where $\rho$, ${\bf v}(t,{\bf x})$, $p(t,{\bf x})$ are the density, the velocity, and the pressure fields, $\eta \equiv \rho \nu$ is the dynamic viscosity, and $\nu$ is the fluid's kinematic viscosity.

The time-dependence of the velocity field gives rise to hydrodynamic memory effect, where the motion of a solid body immersed in viscous fluid, as well as its interaction with another solid body inside the fluid, is affected by the past history of its motion~\cite{bous,basset,oseen,1sp,hyd4,onesph,2sp,hyd6,2spnew,ourjcp}. However,  timescales over which hydrodynamic memory is relevant are dictated by $\tau_{\rm hyd} = \frac{a^2}{\nu}$ where $a$ is the size of the object, being $\sim 1\ \mu{\rm s}$ in the current experiment, and  $\nu \sim 10^{-6}\ {\rm m^2/s}$ in the current experiment where we use water. Overall, this gives $\tau_{\rm hyd} \sim 1\ \mu{\rm s}$ which is much smaller than the time-resolution of the current experiment where we observe bead movement on $\rm ms$ time scales. For this reason we neglect the effect of hydrodynamic memory, which amounts to neglecting the time-derivative in Eq.(\ref{ulns}), leading to the steady linear Navier-Stokes equation~\cite{LL, micro}:
\begin{equation}
 {\nabla p} - \eta\nabla^2 {\bf v}=0\label{sns}.
\end{equation}
Eq.(\ref{sns}) is then the time-dependent statement of momentum conservation,
\begin{equation}
    \nabla \cdot {\bf T} = {\bf 0}
\end{equation}
where ${\bf T} \equiv \rho p {\bf I} - \eta \nabla {\bf v}$ is the stress tensor for the incompressible Newtonian fluid, with ${\bf I}$ defining the identity operator. When an external force field ${\bf f}({\bf x})$ acts upon the fluid, the equation is modified such that
\begin{equation}
{\nabla p} - \eta \nabla^2 {\bf v}={\bf f}.\label{sns2}
\end{equation}

\begin{figure}
\includegraphics[width=\textwidth]{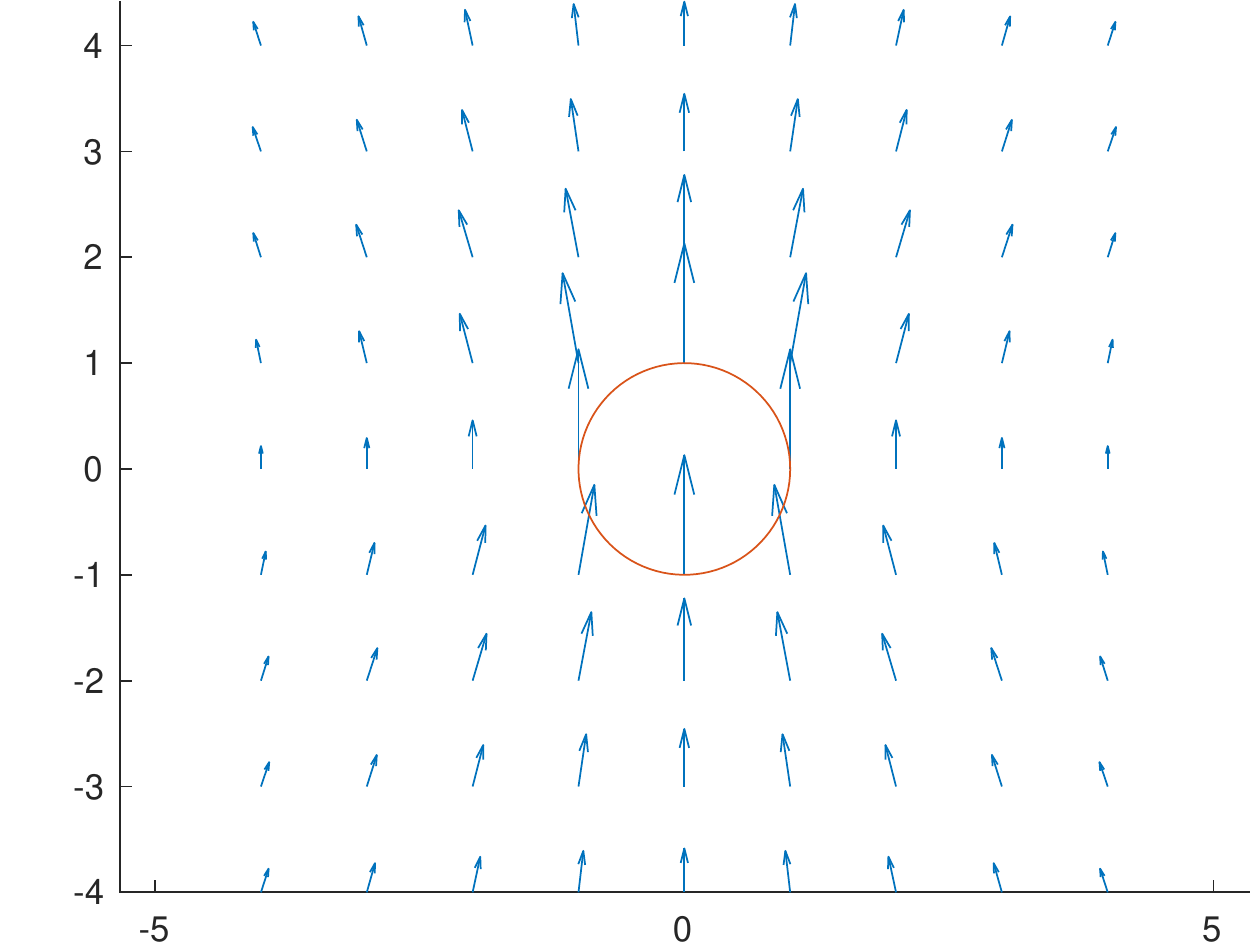}
\caption{The velocity field  {\bf v} ({\bf x}) outside a non-rotating sphere moving with a constant velocity. This vector field is approximately proportional to the force exerted on a stationary sphere centered at each location.}
\label{onefig} 
\end{figure}

When a solid object such as a microbead moves within viscous fluid,
the fields ${\bf v} ({\bf x})$ and $p({\bf x})$ outside the object can be obtained by solving Eq.(\ref{sns}) with appropriate boundary condition at the object's surface. The no-slip boundary condition is usually employed, where the tangential component of the fluid velocity as well as the normal component matches that of the solid body. Under no-slip boundary condition, the velocity field outside the rigid sphere of radius $a$ centered at ${\bf x} = {\bf 0}$, moving with translational velocity $\bf U$ and rotating with the angular velocity ${\boldsymbol \omega}$, is given by~\cite{LL,micro}
\begin{equation}
    {\bf v} ({\bf x}) = \left(\frac{3a}{4r}+\frac{a^3}{4r^3}\right) {\bf U} +  \left(\frac{3a}{4r}-\frac{3a^3}{4r^3}\right)  ({\bf e}_r \cdot {\bf U}) {\bf e}_r + {\boldsymbol \omega} \times {\bf x} \frac{a^3}{r^3}\quad (|{\bf x}| > a),  \label{stksol}
\end{equation}
where $r \equiv |x|$ and ${\bf e}_r \equiv {\bf x}/{r}$. The velocity field {\bf v} ({\bf x}) in Eq.(\ref{stksol}) for $\boldsymbol \omega = \bf 0$, inside the plane containing $\bf U$, is shown in Figure \ref{onefig}. 

The force ${\bf F}$ and the torque ${\boldsymbol \tau}$ exerted by the fluid on the solid body is obtained by integrating the moments of the fluid stress tensor along the object's surface object. For a spherical object of radius $a$ centered at ${\bf x} = {\bf 0}$ inside a fluid with ambient velocity field ${\bf v}^\infty({\bf x})$, moving with translational velocity $\bf U$ and rotating with angular velocity $\boldsymbol \omega$, the result is expressed by Fax\'{e}n's laws~\cite{micro}:
\begin{eqnarray}
{\bf F} &=& 6 \pi \eta a \left(1 + \frac{a^2}{6} \nabla^2 \right) {\bf v}^\infty ({\bf x}) |_{\bf x=0} - 6 \pi \eta a {\bf U}, \nonumber\\
{\boldsymbol \tau} &=& 8 \pi \eta a^3 \left(\nabla \times {\bf v}^\infty({\bf x})|_{\bf x = 0} - \boldsymbol{\omega}\right)\label{fax}
\end{eqnarray}
where the ambient velocity field satisfies the following properties:
(i) ${\bf v}^\infty ({\bf x})$ matches ${\bf v} ({\bf x})$ at infinity, such that $\lim_{|{\bf x}| \to \infty} |{\bf v} ({\bf x}) - {\bf v}^\infty ({\bf x}) | = 0$;
(ii) The force field ${\bf f}({\bf x}) = \nabla \cdot {\bf T}$ computed from ${\bf v}$ and ${\bf v^\infty}$, should be the same outside the sphere; 
(iii) ${\bf v^\infty}$ satisfies Eq.(\ref{sns}) inside the sphere as well as at the surface, by contrast to ${\bf v}$.  In particular, ${\bf v^\infty}$ does not satisfy the boundary condition at the sphere's surface. For a sphere moving with translational velocity ${\bf U}$ and angular velocity ${\boldsymbol \omega}$ inside the quiescent fluid, ${\bf v}^\infty ({\bf x}) = \bf 0$ and Eq.(\ref{fax}) reduces to
 \begin{eqnarray}
{\bf F} &=&  - 6 \pi \eta {\bf U}, \nonumber\\
{\boldsymbol \tau} &=& -8 \pi \eta a^3 \boldsymbol{\omega},
\end{eqnarray}
recovering expressions for the Stokes drag. 

\begin{figure}
\includegraphics[width=\textwidth]{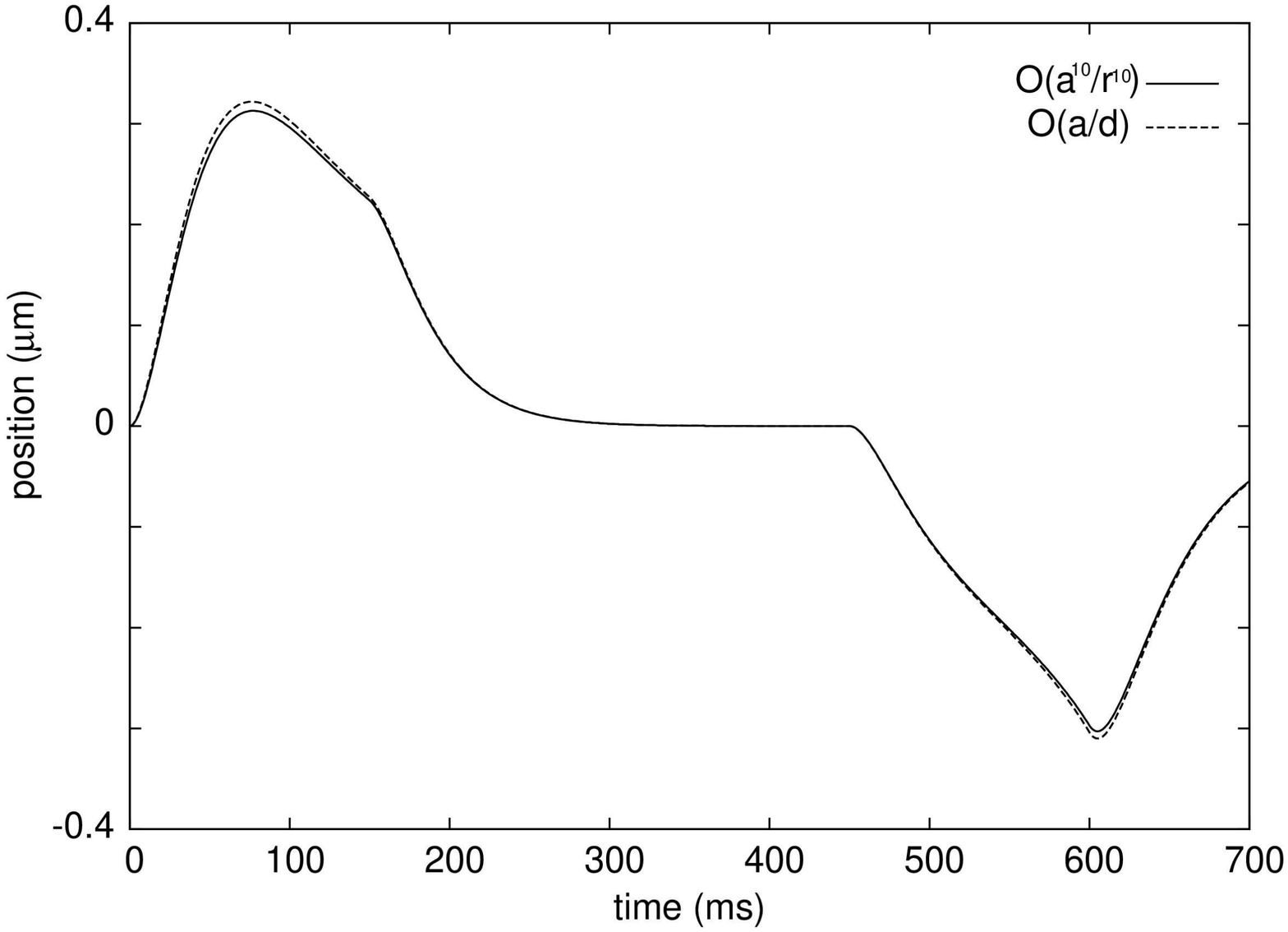}
\caption{The comparison of the computational result for the linear scan, where up to $O(\frac{a^{10}}{d^{10}})$ terms were used in one computation (solid line) and only up to $O(\frac{a}{d})$ terms were used in the other.}
\label{comp} 
\end{figure}

If there is another sphere $B$ centered at ${\bf x} = {\bf d}$, in the presence of the first sphere $A$ at ${\bf x} = {\bf 0}$ moving with translational velocity ${\bf U}$ and rotating with angular velocity $\boldsymbol \omega$, Eq.(\ref{fax}) is invoked again at ${\bf x} = {\bf d}$ to compute the force exerted on $B$, but this time the velocity field given by Eq.(\ref{stksol}) becomes the approximate ambient field, by extending the fluid velocity field in Eq.(\ref{stksol}) to region inside the sphere $A$, with nonzero force field ${\bf f}({\bf x})$ at $|{\bf x}|=a$~\cite{gfax,nsp1,nsp2}, so that it is defined in all regions outside of the sphere $B$.  The reason that such velocity field is not exactly the ambient field for $B$ is on account of the presence of the sphere $B$ which necessarily perturbs the velocity field and changes the force field at $|{\bf x}|=a$, violating the condition (ii) above for the ambient field. This is connected with the fact that the actual velocity field at the surface of $B$ can be obtained only iteratively: The velocity field in Eq.(\ref{stksol}), now written as ${\bf v}_A({\bf x})$, does not satisfy the boundary condition at the surface of $B$. Once we add the necessary correction, denoted as ${\bf v}_{AB}$, then the boundary condition at $A$ is violated. We then continue adding the correction terms. Each correction to the velocity field at $B$ or the force exerted on $B$ is on the order $O({a^2}/{d^2})$ compared to the previous one, assuming the radius of the sphere B is also of order $a$. Therefore, ${\bf v}_A({\bf x})$ is the ambient field for $B$ only if we neglect the error of $O({a^3}/{d^3})$. Also, the $\nabla^2 {\bf v_A}$ term in the expression for ${\bf F}$ in  Eq.(\ref{fax}) is at most of $O({a^3}/{d^3})$ at $B$. Therefore ${\bf v}_A({\bf x})$ is approximately the force experienced by a stationary sphere whose center is located at $\bf x$, if we neglect the error of order $O({a^3}/{r^3})$. In fact, the result of the two-sphere computation where  $O({a^3}/{d^3})$ terms are discarded and only $O({a}/{d})$ terms are kept, shows qualitatively the same trajectory as the more exact results,  as shown in Figure \ref{comp} where the computational result  for the linear scan using up to $O({a}/{d})$ terms is compared with the one where up to $O({a^{10}}/{d^{10}})$ terms are kept. The results for two-body and three-body shear also show the same behavior (data now shown). Therefore, we may safely say that Figure \ref{onefig} displays the force felt by a stationary sphere centered at each position. Obviously, such a sphere is dragged in the direction of ${\bf U}$. When the sphere $B$ moves with the velocity ${\bf U}_B$, then it feels additional drag force $-6 \pi \eta a {\bf U}_B$. When the magnitude of ${\bf U}_B$ matches that of the ambient fluid velocity, the sphere can coast without any force exerted by the fluid. All discussions above can be applied to the torque and the angular velocity as well.

\begin{figure}
\includegraphics[width=\textwidth]{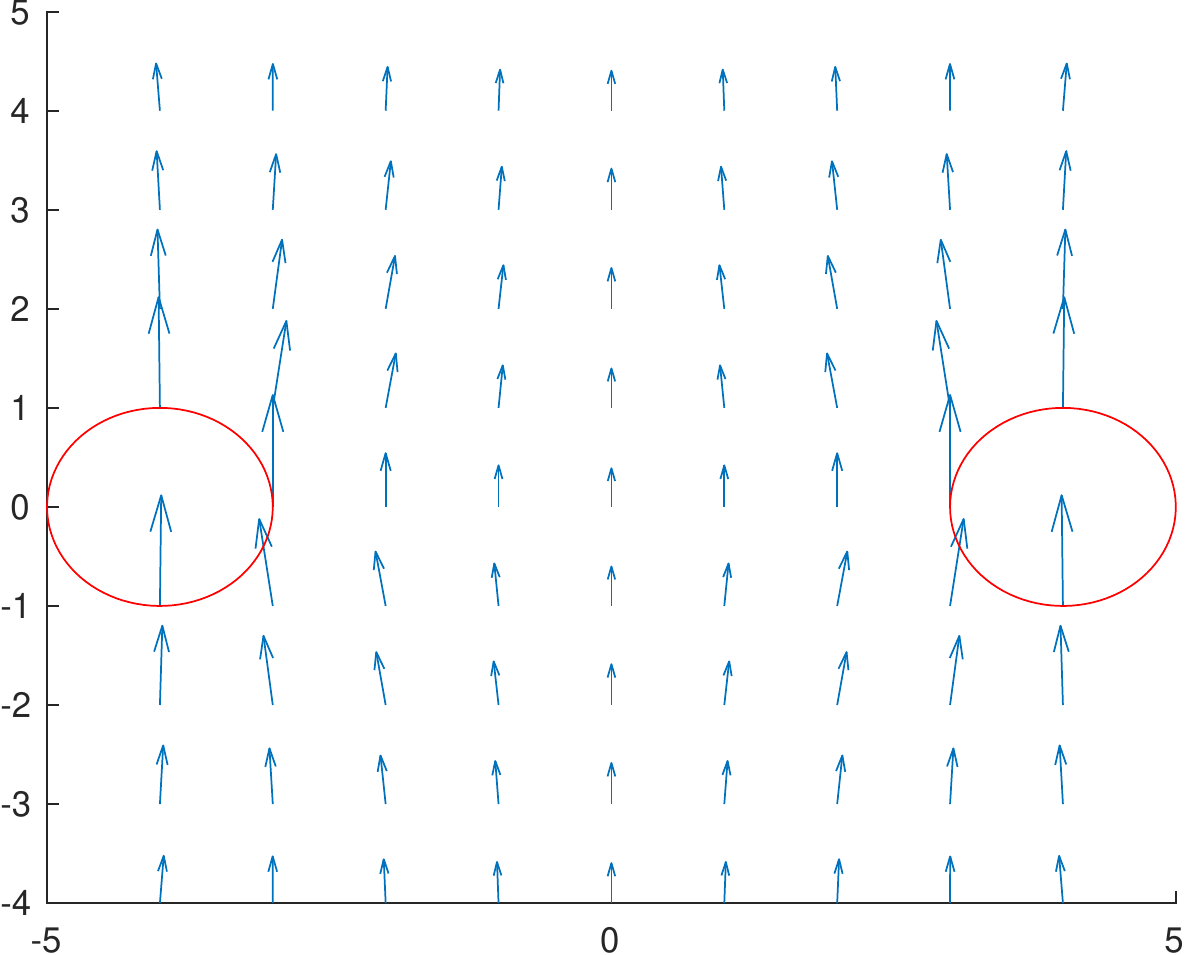}
\caption{The velocity field  {\bf v} ({\bf x}) outside two non-rotating spheres moving with a constant velocity. This vector field is approximately proportional to the force exerted on a stationary sphere centered at each location.}
\label{twofig} 
\end{figure}

The superposition of velocity fields of two spheres of equal size is shown in figure~\ref{twofig}. The error compared to the exact velocity field of the two sphere is again  $O(a^3/r^3)$. Again, it is proportional to the force felt by a third sphere centered at each position, with the error of $O(a^3/r^3)$.

The force ${\bf F}$ exerted by the fluid on the spheres and the velocities ${\bf U}$ of the spheres are related by the friction tensor
\begin{equation}
    {\bf F} = -\boldsymbol{ \zeta} {\bf U} \label{ze}
\end{equation}
where now the indices for the spheres and the spaces are implicit, so that ${\bf F}$ and ${\bf U}$ are $3N$-dimensional vectors, and $\boldsymbol{ \zeta}$ is a $3N \times 3N$ tensor, for $N$ spheres. Inverting Eq.(\ref{ze}), we get
\begin{equation}
    {\bf U} = -\boldsymbol {\mu} {\bf F} \label{mu}
\end{equation}
where $\boldsymbol {\mu}$ is the inverse of $\boldsymbol{ \zeta}$, called the mobility tensor. The mobility tensor for $N$ sphere has been computed to order $O(a^7/d^7)$~\cite{nsp2}, which we use in the current work. In particular, the mobility tensor for the three-body case is given by Eq.(2) in the main text. The mobility tensor for the two-body case is easily obtained by taking the limit of $R_{AC}, R_{BC} \to \infty$, in Eq.(2) of the main text, which is
\begin{eqnarray}
\tilde {\boldsymbol \mu}_{AA}  &=& \frac{1}{a} - \frac{15 b^3}{4 R_{AB}^4} {\bf \hat e}_{AB} {\bf \hat e}_{AB}  - \frac{15 c^3}{4 R_{AB}^4} {\bf \hat e}_{AC} {\bf \hat e}_{AC} + O(\epsilon^{6}),  \nonumber\\
\tilde {\boldsymbol \mu}_{AB}  &=& \frac{3}{4R_{AB}}({\bf I}+{\bf \hat e}_{AB}{\bf \hat  e}_{AB}) + \frac{a^2+b^2}{4 R_{AB}^3}({\bf I}-3 {\bf \hat e}_{AB}{\bf \hat  e}_{AB}) \nonumber\\
&&+O(\epsilon^6).
\end{eqnarray}

When there are external forces ${\bf F}_{\rm ext}$ acting on the spheres, then the net force ${\bf F}_{\rm net} = {\bf F}_{\rm ext} + {\bf F} $ makes the spherical bead accelerate until it reaches the terminal velocity and ${\bf F}_{\rm net}$ becomes zero. This happens within a time scale of $\tau_B \equiv \frac{a^2}{9 \nu}(1+2 \frac{\rho_s}{\rho_f})$ where $\rho_s$ and $\rho_f$ are the densities of the spherical bead and the fluid, respectively. In our case $\tau_B \sim 1\ \mu {\rm s}$, being much less than the observational time scale of $\sim {\rm ms}$ and with no appreciable change of external forces during that time. Therefore, the current experiment is done in the overdamped regime where we may assume ${\bf F}_{\rm ext} = - {\bf F}$ all the time. Then Eq.(\ref{mu}) becomes 
\begin{equation}
    {\bf U} = {\boldsymbol \mu} {\bf F_{\rm ext}} 
\end{equation}
where the position-dependent external force $\bf F_{\rm ext}({\rm x})$ is supplied by the optical trap. We then numerically integrate this equation, with the time step of $\Delta t = 1 \mu {\rm s}$, to get the computational result displayed in the main text.
 
\clearpage

\section{Details on experiments}

\begin{figure}[b]
\includegraphics[width=\textwidth]{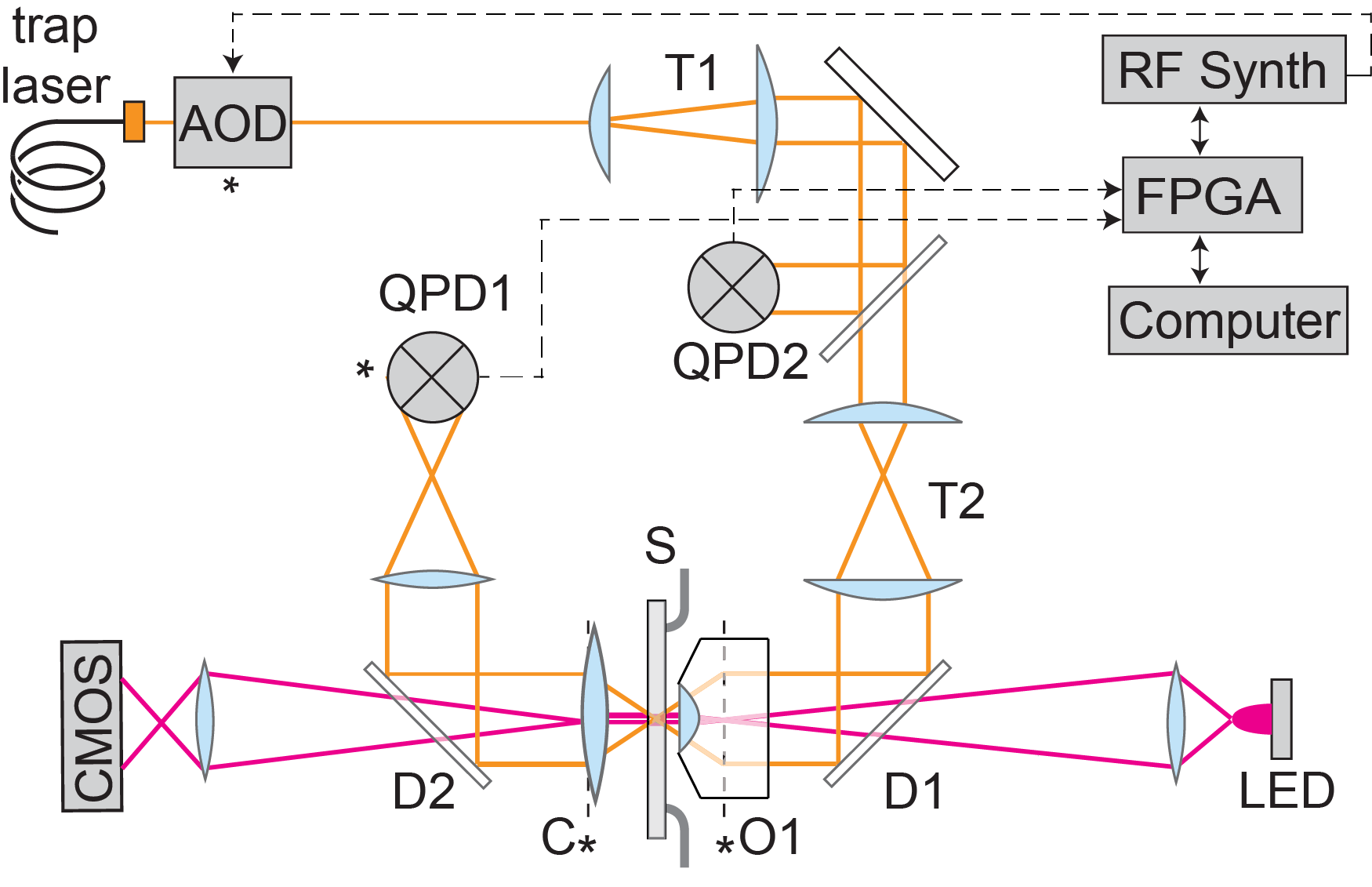}
\caption{Schematic layout of dual-trap optical tweezers instrument used in this investigation. Orange and red lines indicate trap laser and visible imaging illumination paths respectively. The trap laser passes through an acousto optic deflector (AOD) device, which rapidly ($\sim$ 100 kHz) deflects the laser in two-dimensions cyclically between three independently controlled deflection angles corresponding to three trap locations ({\it i.e.}, the probe bead trap and two flanking scan bead traps). The trap laser is then expanded and imaged by a pair of telescopes (T1 and T2) onto the back aperture of the objective lens (O1) which focuses the laser within a sample chamber (S) to form the traps. $\sim$ 5\% of the laser is picked off and sent to QPD2 for intensity monitoring to provide stabilizing feedback on the trap intensity. A subsequent condenser lens (C) collects trap laser light exiting the sample chamber and passes it to a bead position detector (QPD1). We note that, in general, bead position measurements were made via directly imaging the bead motions using the CMOS camera. Electrical control and measurement are indicated by black dashed lines, while a field programmable gate array (FPGA) chip is responsible for coordinating all measurement and control including the direct digital control of the direct digitial synthesis (DDS) RF source controlling the AOD. In addition, the FPGA synchronized camera data acquisition with trap scan initiation. D1 and D2 are dichroic mirrors.}
\label{suppfig4_tweezer_layout} 
\end{figure}

\begin{figure}
\includegraphics{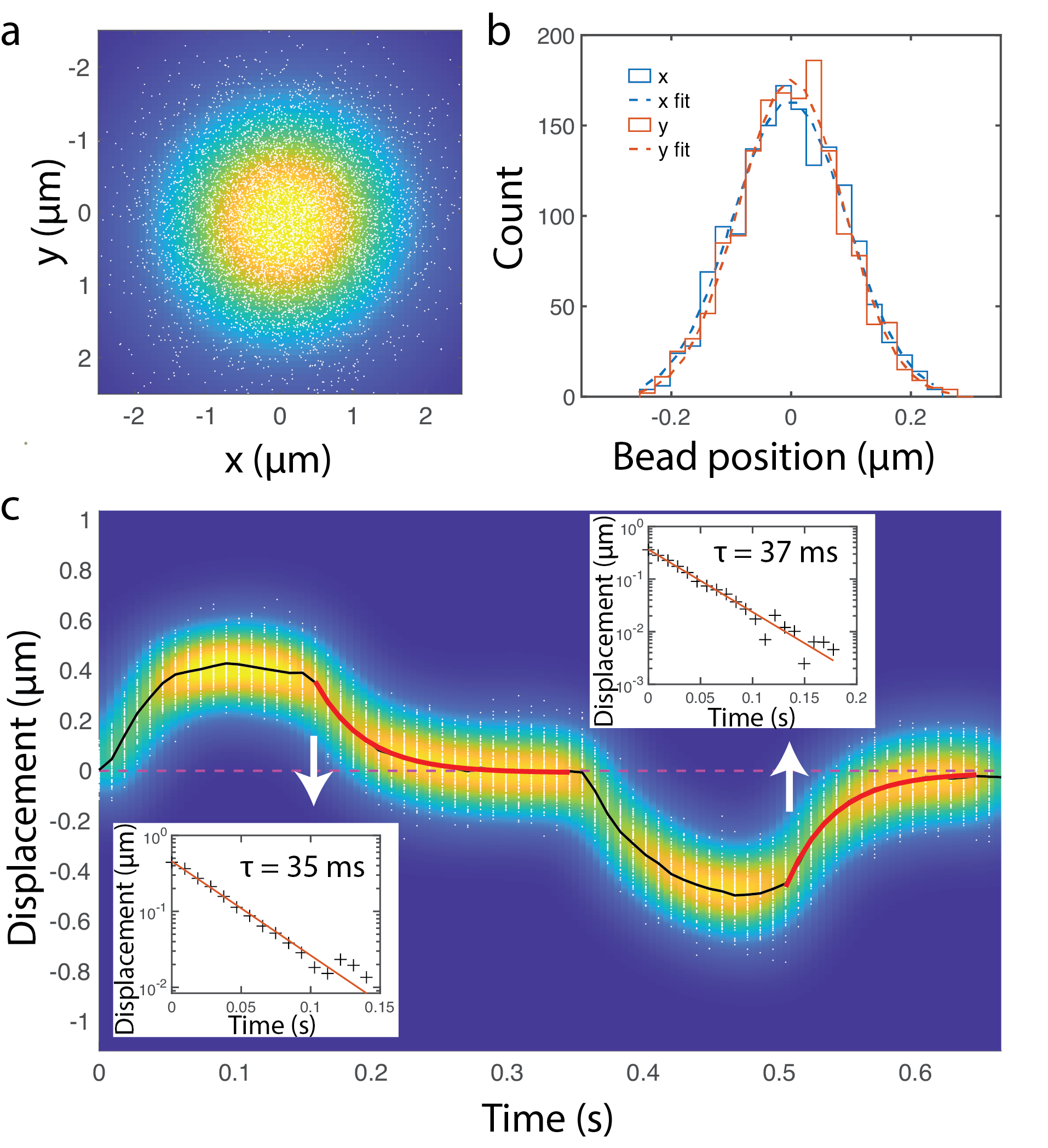}
\caption{Characterization and calibration of a typical optical trap used in the experiment. The exact trap stiffness varied slightly for different experiments. a)  Plot of a trapped bead's position within a fixed location trap ({\it i.e.}, no scanning). Each dot is the bead's location in one image frame from a sequence of frames of duration equivalent to a set of 100 scans (as in the actual scan experiments). The traps have highly symmetric Gaussian profiles. b) Example distribution of trapped bead positions in x and y during the final pause intervals at the ends of a set of 100 scans for a trapped bead (the final 5 frames of each scan). The dashed lines are Gaussian fits. The root-mean-square (RMS) of the data (not the fit) is used to determine the trap stiffness $k$ from the standard relation $\rm{RMS} = \sqrt{k_BT/k}$, where $k_B$ is the Boltzmann constant and $T$ the temperature. c) Main: example probe bead motion for a combined set of 100 two-scan-bead shear scans. The dots are the individual probe bead positions in each image frame from all scans while the black line is the average probe bead displacement over all scans. The red curves are simple single exponential fits to the relaxation of the probe bead in the trap upon pausing of the scan before changing direction (first fit) and at the end of the scan (second fit). Insets are zooms of the same decay data and fits with log-log scales with the decay time constant $\tau$ indicated. Additional control experiments were also performed (not shown) where for a single bead held in a trap, the trap is instantaneously displaced (the AOD trap steering method allows for this) and the bead subsequently relaxes to the new trap location. In that case the same bead relaxation time constants were also observed. The bead relaxation time constant is simply the bead viscous relaxation time constant $\tau = 6\pi\rho\nu R/k$ where $\rho = 997 kg/m^3$ is the solution (water) density, $\nu = 9.381 m^2/s$ is the solution dynamic viscosity at the experiment temperature $T = 22.8 \rm{C}$, and $R = 1.08 \mu m$ is the bead radius. For all beads and traps used in these investigations, we performed measurements of the individual bead RMS in a fixed trap and viscous relaxation time constants measured both during scan pauses and in a follow-up trap jump experiment for the same bead, computed the corresponding trap stiffness for all three methods, and found that all three methods agreed to within 5\%.}
\label{suppfig5_trap_calibration} 
\end{figure}

\begin{figure}
\includegraphics[width=\textwidth]{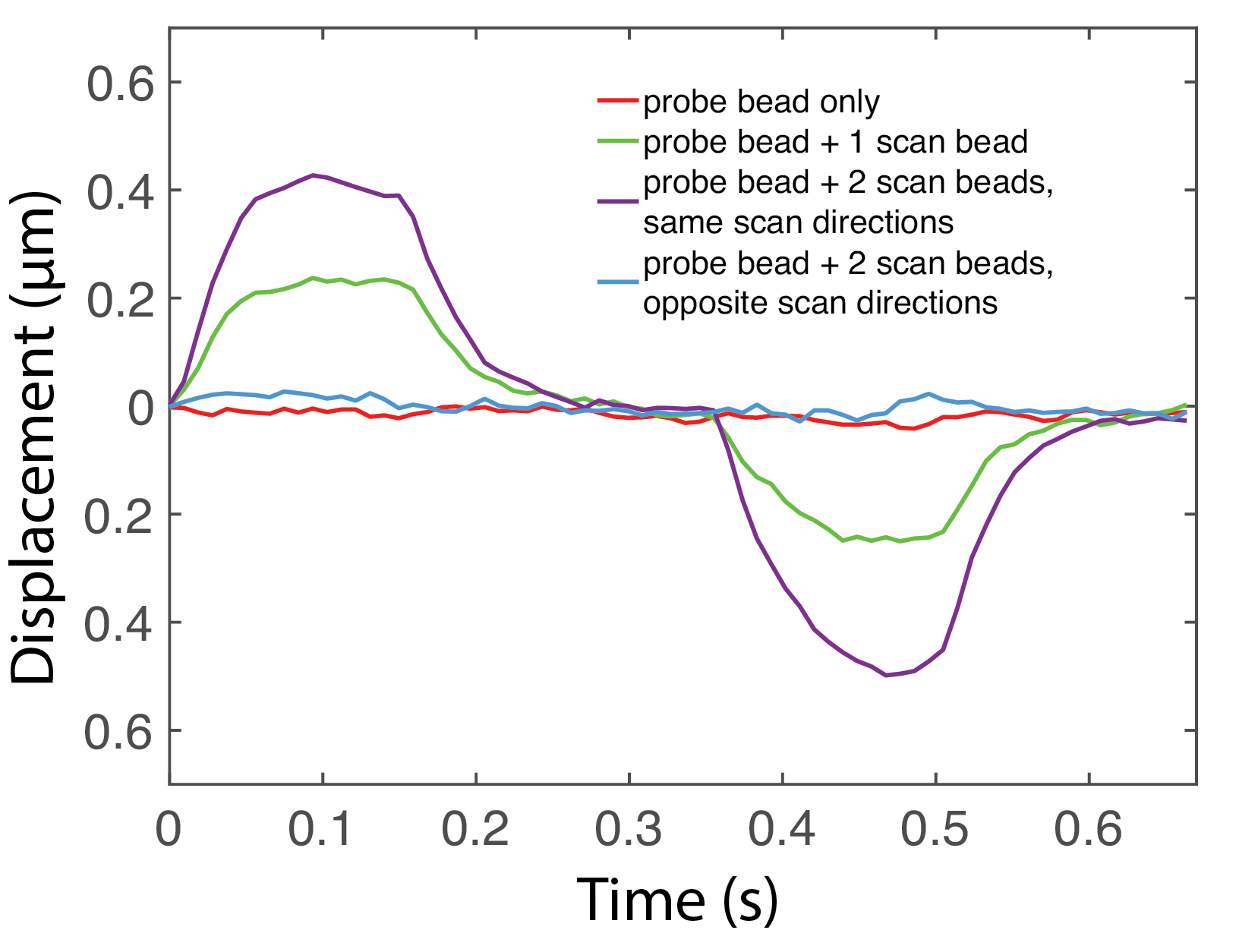}
\caption{Probe bead displacement vs time for shear mode scanning experiments as in Fig. 2 including control scans. The plots are the mean probe bead displacement for 100 scans and are the same bead sets and data as shown in part in Fig. 2. The initial configuration is as in Fig. 2a with either zero (``probe bead only"), one or two scan beads in the pair of traps flanking the central probe bead. For no scan beads in the scanning traps, no displacement of the probe bead is observed, as expected and verifying that the scanning traps themselves are not introducing a net force on the probe bead. The mean probe bead displacements for one or two scan beads are replicated from Fig. 2b and c respectively. As an additional control, the two scan beads are also scanned in opposite directions, {\it i.e.}, one in +y and the other in the -y direction (as shown in more detail in the following Supp. Fig. \ref{suppfig7_shear_controls2}). As expected, the forces from the two scan beads on the probe bead cancel and there is no probe bead displacement. As in Fig. 2, each plot is the average of 100 repeated scans over a scan range of $6 \mu {\rm m}$ with a scan speed of $40 \mu {\rm m/s}$.}
\label{suppfig6_shear_controls1} 
\end{figure}

\begin{figure}
\includegraphics[width=\textwidth]{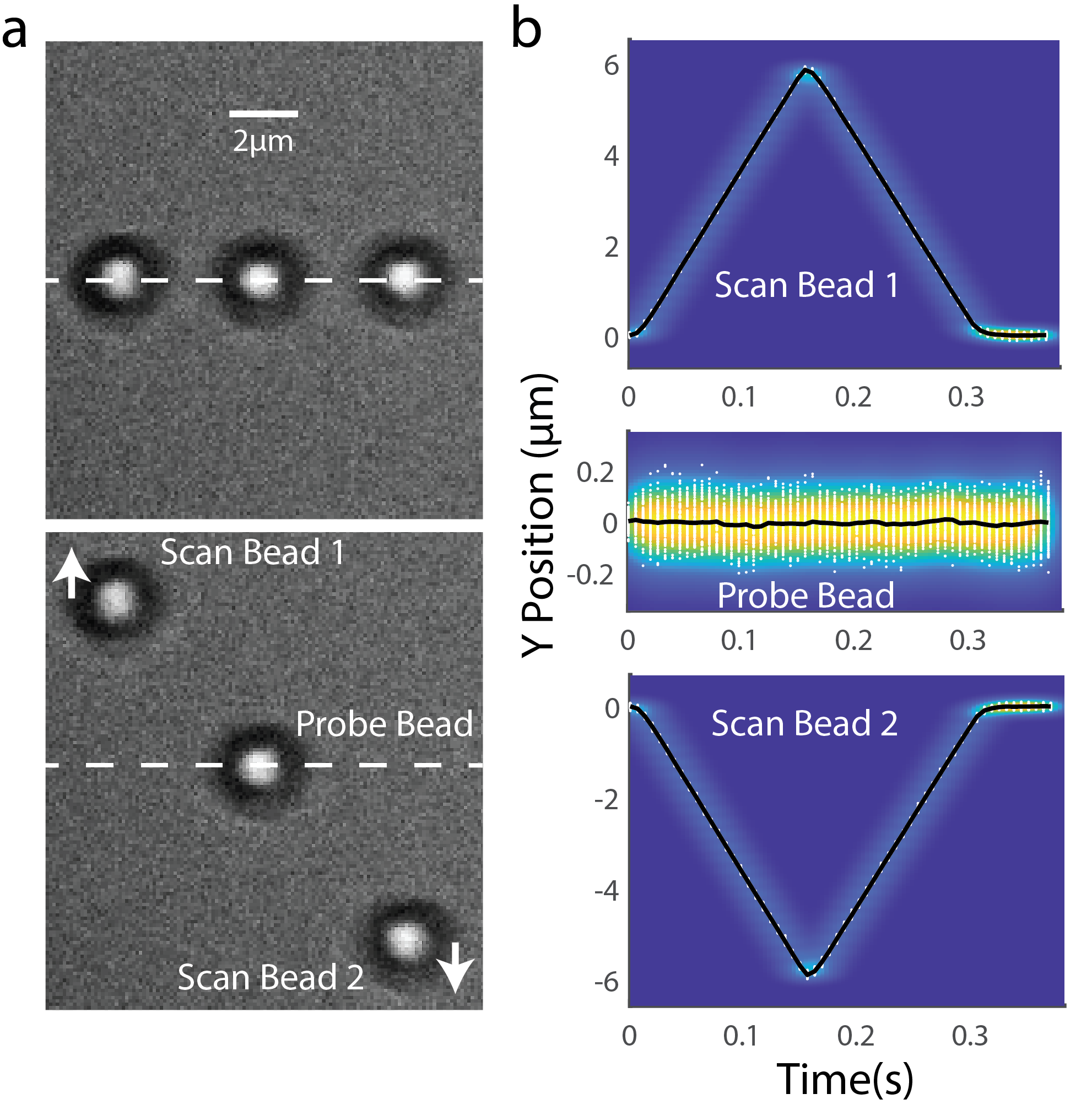}
\caption{Three-bead opposing shear scan control showing no induced probe bead displacement. a) Example images of a three-bead opposing shear mode scanning experiment. Upper: initial arrangement of three beads in a line with two scanning bead flanking a probe bead. The dashed line indicates both the probe bead's fixed trap position as well as the initial scan bead trap positions. Lower: a single image during the scan showing the two scan traps and bead moving in opposite directions (white arrows). b) Opposing shear scan bead positions for a single set of beads averaged over 100 scans (black lines) and individual positions (white dots). The background color map is the displacement distribution. Upper and lower: left and right scan beads respectively. Center: probe bead.}
\label{suppfig7_shear_controls2} 
\end{figure}

\begin{figure}
\includegraphics[width=\textwidth]{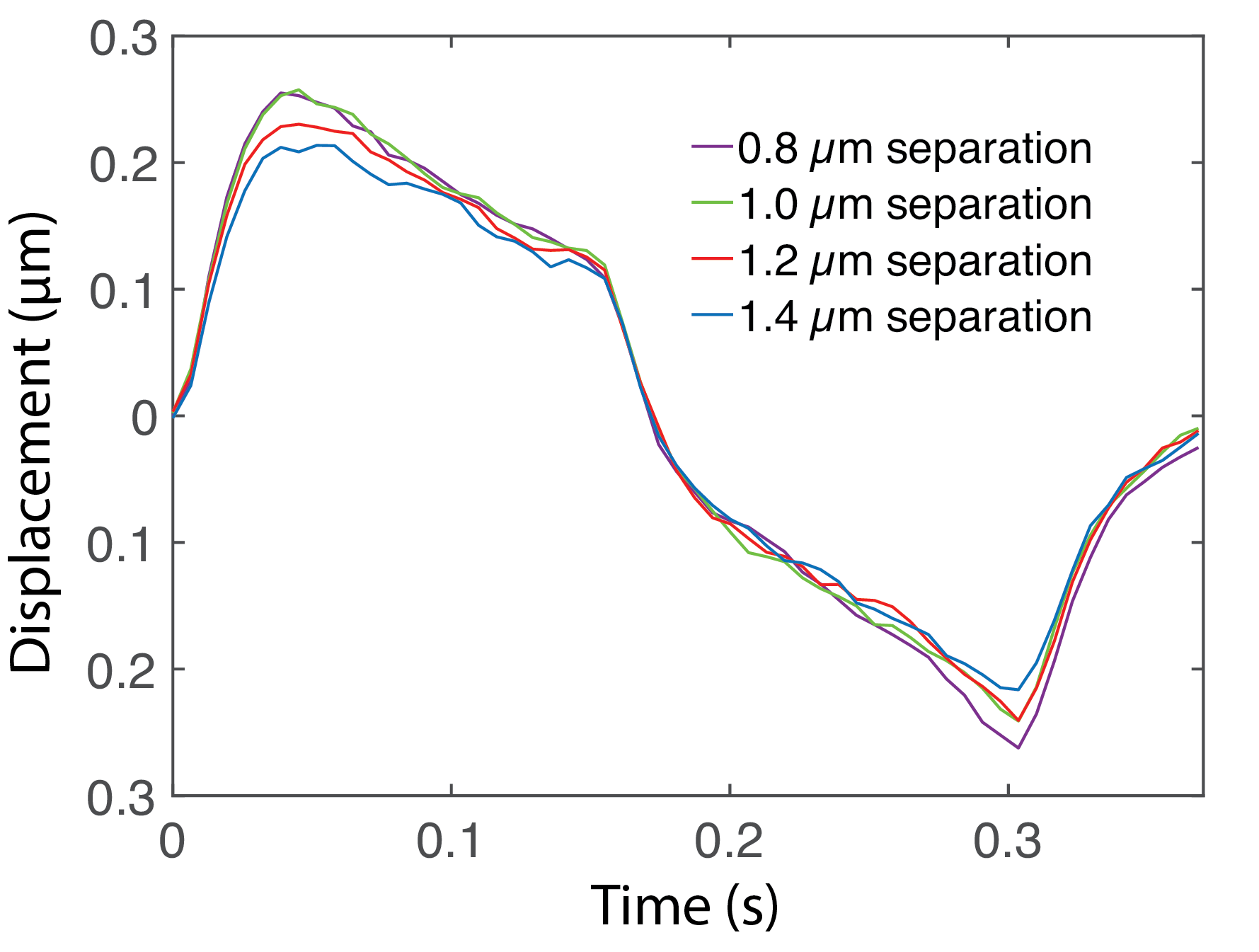}
\caption{Average displacement of probe bead during two bead linear scan showing a small reduction in the maximum probe bead displacement with increasing the initial and final bead separations. Scan protocol is as in Fig. 1 except the scan direction reverses immediately without pause at t = 150 ms). The scan range is $6 \mu {\rm m}$ and the scan speed is $40 \mu {\rm m/s}$. Data are for a single bead set, and each trace is the average of 100 repeated scans.}
\label{suppfig8_linear_separataion_dep} 
\end{figure}

\clearpage

\bibliography{spheres}